\documentclass[aps,prb,twocolumn,superscriptaddress]{revtex4-1}
\usepackage{color}
\usepackage{xcolor}
\usepackage{amsmath}
\usepackage{amssymb}
\usepackage{amsfonts}
\usepackage{graphicx}
\usepackage{physics}
\usepackage{wasysym}
\usepackage[normalem]{ulem}

\def\vk{{\bf k}}
\def\vq{{\bf q}}
\def\vr{{\bf r}}
\def\vR{{\bf R}}
\def\vd{{\bf d}_{l}}
\def\vdp{{\bf d}_{l+1}}
\def\vdm{{\bf d}_{l-1}}
\def\va{{\bf a}_{l}}
\def\vap{{\bf a}_{l+1}}
\def\vam{{\bf a}_{l-1}}
\def\vD{{\bf D}_{l}}

\begin{document}
\title{Slave-rotor theory on magic-angle twisted bilayer graphene}
\author{Shin-Ming Huang}
\affiliation {Department of Physics, Center of Crystal Research, National Sun Yat-sen University, Kaohsiung 80424, Taiwan}
\author{Yi-Ping Huang}
\affiliation {Condensed Matter Theory Group, Paul Scherrer Institute, CH-5232 Villigen PSI, Switzerland}
\author{Ting-Kuo Lee}
\affiliation {Department of Physics, Center of Crystal Research, National Sun Yat-sen University, Kaohsiung 80424, Taiwan}

\begin{abstract}
We investigate the correlated electrons in
the magic-angle twisted bilayer graphene by using the slave-rotor mean-field theory. 
Owing to the extended figure of Wannier orbitals, we study the two-orbital cluster
Hubbard model with spin-valley fourfold degeneracy, focusing around half filling of
valence bands below the neutrality point.
The theory predicts multiple Mott insulator phases at fractional fillings not only for integer
charges per moir\'e site, and it demonstrates that long-range electron hopping is highly suppressed because multiple-charge excitations are induced. 
Furthermore, the Kekul{\'e} valence bond order is investigated and is found to extend the Mott insulator phases to
occupy a finite doping region.
Adjacent to Mott insulator phases, superconducting domes emerges by virtue of spin-valley fluctuations.
This work has provided a primal understanding and interesting phenomena of the correlated system, and for its novel interaction the model might produce plenty of possibilities waiting to be explored.

\end{abstract}
\date{\today}
\maketitle
\section{Introduction}
Recently twisted bilayer graphene (TBG) has aroused considerable interest since the
discovery of strongly correlated insulating states and
superconductivity~\cite{cao2018unconventional,cao2018correlated}. 
It is quite surprising that strongly-correlated phenomena could emerge from the carbon atoms
which have weak interaction effect comparing with strongly-correlated transition metal
ions.
%that leads to various unconventional superconductivity such as cuprates and
%heavy-fermion superconductors.
%
The successful fabrication of TBG and its novel phenomena opens another route of
promising applications in van der Waals heterostructures~\cite{geim2013van}.
Owing to a weak van der Waals force tying them, two graphene flakes are manipulable to
stack with a relative twist angle, giving rise to a long-period moir\'e superlattice.
The moir\'e lattice parametrized by the twist angle infers a tendency of moir\'e band
narrowing, and theory predicts extremely flat bands can emerge at magic
angles~\cite{Bistritzer2011,LopesDosSantos2007,Shallcross2010,Trambly2012,Tarnopolsky2019}.
At the magic angle $\theta^{\star}\approx 1.05$, the moir\'e lattice constant is $a= 13$
nm, and the charge density required to fill a moir\'e band (including degeneracy from
spin and valley) is as low as $n_s\approx 2.6 \times 10^{12}~\text{cm}^{-2}$.
The low-energy bands have a bandwidth of the order of 10 meV~\cite{cao2018correlated},
serving as an unstable ground against electron-electron interaction.

A great advantage in TBG to investigate the electronic structure is the controllable
carrier concentration that experiment can access major part of the phase diagram simply
by electronic gating due to a low density of the moir\'e band.
In transport studies, correlated gaps at partial band fillings have been experimentally
observed~\cite{cao2018unconventional,cao2018correlated,Xie2019,choi2019electronic,kerelsky2019maximized,polshyn2019large,Lu2019}.
With doping, superconductivity appears and resides on either side of the correlated
insulating phase as superconducting domes in the temperature-density phase diagram,
reminiscent of strongly correlated high-$T_c$ superconductors~\cite{Lee2006}.
In addition, the quantum oscillations display vanishing of Fermi surfaces and an
increase of the effective mass when the correlated insulating phase is approached,
signifying a doped Mott insulator~\cite{cao2018unconventional,cao2018correlated}.
Through analyzing temperature behavior as well as the magnetic response in
conductance~\cite{cao2018correlated}, the Mott gap (correlated gap) was estimated
to be 0.3 meV. 
In comparison with the bandwidth, the small Mott gap reveals that the system might be in close proximity to the Mott transition point. 
Although there are many features corresponding to Mott physics, some observations
conflict our understanding: the phase diagram is quite asymmetric on two doping sides of
the Mott insulator phase and the Landau fan suggests broken spin-valley
symmetry~\cite{cao2018unconventional,cao2018correlated,Lu2019}. 
Nevertheless, TBG as an unconventional superconductor is evident for its relative high
critical temperature $T_{c}\sim 2 \text{~K}$~\cite{cao2018unconventional,Lu2019}.
To dates, most theoretical studies on superconductivity belong to the weak coupling
theories~\cite{guo2018pairing,thomson2018triangular,peltonen2018mean,liu2018chiral,gonzalez2019kohn,tang2019spin,roy2019unconventional,ray2019wannier,lian2019twisted,You2019},
and a theory in an intermediate-coupling or a strong-coupling regime is
requisite~\cite{dodaro2018phases,Ochi2018,xu2018kekule,kennes2018strong,classen2019competing,thomson2018triangular}.

In this work, we investigate TBG as a doped Mott insulator by the cluster
Hubbard model proposed in Refs.~\onlinecite{Po2018,Koshino2018}.
The theory is developed based on the peculiar extended Wannier
orbitals~\cite{Po2018,Koshino2018} where the electron repulsion is a cluster charge
(total charge in a hexagon) interaction.
Studying such a model with unconventional interaction is challenging because
of the non-trivial low-energy constraint on a cluster. To conquer the difficulty, we use the slave-rotor approach to deal with the correlation effects. 
The strategy has been used widely in the study of correlated strongly spin-orbit coupled
materials~\cite{pesin2010mott,ko2011magnetism}, but the difference in our method is that the rotor in this work is to depict total charge in a hexagon instead of on a single site. Utilizing the slave-rotor mean-field theory and comparing different approximations, we can simplify the problem and discovery a number of interesting phenomena in the model. Distance-dependent renormalization effects on hopping appear plainly due to the cluster interaction. Therefore, a short-range hopping model becomes sufficient in the strong coupling regime so that the critical value of interaction for the Mott transition becomes less sensitive to the details of the bare band structure. Most notably, the Mott insulator phase appears, if happens, not only at integer fillings (charges per moir\'e) but also at fractional ones as $2\pm1/3$ and so on. We also found the system having a tendency of the Kekul\'e valence bond order (KVB)~\cite{Chamon2000,xu2018kekule} thanks to the slave-rotor representation. The KVB is found to be compatible with the Mott insulator phase, exhibiting a Kekul\'e valence bond solid. Considering spin-valley fluctuations, the model can realize superconducting pairing and gives a phase diagram similar to experimental observations. 

The paper is organized as follows. Section \ref{sec:theory} shows the Hamiltonian. 
Section \ref{sec:slaverotor} demonstrates the slave-rotor theory and discusses the Mott transition. Three approximations are studied: i) the single-site approximation in Sec. \ref{sec:single-site}, ii) two-site approximation in Sec.~\ref{sec:two-site}, and iii) the O(2) nonlinear sigma model in Sec. \ref{sec:NLS}. (Reader who are familiar with the theory and approximations can skip the detail and just refer to the conclusive result in Fig.~\ref{fig:Z}.)
Section \ref{sec:AFMSC} discusses the phase diagram of antiferromagnetic and superconducting orders. In Sec. \ref{sec:kekule}, the Kekul\'e order and its effect to the Mott transition are shown. Finally, we conclude the work in Sec. \ref{sec:conclusion}. Some details of derivation are found in Appendix: the local gauge symmetry in the slave-rotor representation in Appendix \ref{Appendix:localgauge}, and the mean-field Hamiltonian for hopping and the nonlinear sigma model in Appendix \ref{Appendix:MF} and \ref{Appendix:NLS}, respectively. 

\section{Theoretical model} 
\label{sec:theory}
The flat bands of TBG are constructed by the low-energy states at valleys $K$ and
$K^{\prime }$ from two graphene layers.
The two valleys are considered as independent degrees of freedom when the twist angle is small.
Although the electrons from the flat bands in TBG are seen to concentrate on
the AA sites forming a triangular lattice, developing a tight-binding model for
a low-energy effective theory could be subtle. 
One cannot describe the system by an effective tight-binding model on a triangular
lattice since it is unable to produce Dirac points due to symmetry reason\cite{Po2018,Koshino2018,Kang2018}.
Instead, the effective model should be based on Wannier orbitals on honeycomb-lattice sites (BA and AB
sites). 
Theoretical studies report that two Wannier orbitals from BA and AB
sites look like fidget-spinner orbitals that their distributions peak on
three adjacent hexagon centers (AA sites) \cite{Po2018,Koshino2018,Kang2018} [refer to Fig.~\ref{fig:lattice}(b)]. 
The Wannier orbitals, respecting $C_3$ rotational symmetry, show equal amplitude but $\pm 2\pi/3$ phase
differences on three adjacent sites; in other words, two Wannier orbitals
at valley $K$ have angular momentum of $-1$, while the other two at valley $%
K^{\prime }$ take angular momentum of $+1$ because of time-reversal symmetry.

Based on these Wannier orbitals, a hopping Hamiltonian is
written as 
\begin{equation}
H_{t}=\sum_{ij}\sum_{\kappa =\pm} \sum_{\sigma =\uparrow ,\downarrow
}\left(t_{\kappa,ij} - \mu \delta_{ij} \right)c_{\kappa i \sigma }^{\dagger
}c_{\kappa j \sigma },
\end{equation}%
where $\kappa$ and $\sigma$ denote valley and spin states, respectively, and
two sublattices are implicit in the site notation and will be specified by A
and B later. 
Because of time-reversal symmetry, hopping integrals $%
t_{\kappa,ij}=t^{*}_{\bar{\kappa}, ij}$. 
We will adopt the hopping integrals constructed by Koshino et al. subjected to maximally localized Wannier
orbitals at the magic angle $\theta^{\star}$. 
We consider only hopping up to fifth nearest neighbors ($r=\sqrt{3}a$) unless specified otherwise.
For convenience, we will name the n-th nearest neighbor (nNN) hopping integral $t_{n}$ ($n=1, \ldots,5$).
Although it was noted that to capture a better band structure at high energies, long-range hopping
integrals were small but required, we will show in this work that electronic
correlation, giving different order of renormalization, will highly reduce long-range
hopping.

Referring to Fig.~\ref{fig:lattice}(b) and imagining that a lobe of a Wanner orbital takes charge of $e/3$ and concentrates completely at the center of a hexagon, Wannier orbitals can overlap between onsite, 1NN, 2NN, and 3NN sites with numbers of lobes in 3:2:1:1. 
Since Wannier orbitals at the corners of a hexagon extends to the center of the
plaquette, the electrons on the six corners will interact through the wave function
overlap at the center of the plaquette.
In the approximation that the interaction occurs only at the overlapping region, the number of lobes at a plaquette center will determine the interaction at that hexagon.
As a result, the cluster interaction is introduced as
\begin{equation}
H_{U}=U\sum_{i_{c}}\left( Q_{i_{c}}-C\right) ^{2},
\end{equation}%
where $i_{c}$ runs over hexagon sites (the subscript $c$ is for the hexagonal cluster), the hexagon charge $Q_{i_{c}}=\sum_{j
\in i_{c}} n_{j} $ is the particle number of six sites in the hexagon ($%
n_{j}=\sum_{\kappa,\sigma} c_{\kappa j \sigma}^{\dagger }c_{\kappa j \sigma
} $ includes two spins and two valleys), and $C$ is the charge reference. 
In this work, we are interested in the region around the Mott insulator phase
at half filling of valence bands below neutrality, where there is one
electron per lattice site. So we take $C=6$ and call the case undoped.
When a hexagon has five or seven particles, it gets an energy of $U$. 
Our model for the system is 
\begin{equation}
H=H_t+H_{U}.
\end{equation}
It is estimated that the system could enter the strong correlation regime, so that
the low-energy states are confined to a restricted Hilbert space in the $%
U/t\rightarrow \infty $ limit. 
Rather than no doubly occupied site in the Hubbard model, the present interaction, in some sense, demands a weaker
constraint: six particles in a hexagon. 
This leads to high degeneracy \cite{Ochi2018} and high dimensions of the
restricted Hilbert space, in which more configurations are allowed; for instance, a
hexagon can be of no holon (empty site) and no doublon (doubly occupied
site), one, two, or three holon-doublon pairs, not to mention spin and valley configurations. 
Nevertheless, the present one exhibits Mott physics as well, since any particle's movement brings about
changes of hexagon charges nearby (see Fig.~\ref{fig:phase}) that is
unlikely to happen because of an $\mathcal{O}(U)$ energy cost. 

\section{Slave rotor theory of the Mott transition}
\label{sec:slaverotor}
\begin{figure}[tb]
\begin{center}
\includegraphics[width=0.5 \textwidth]{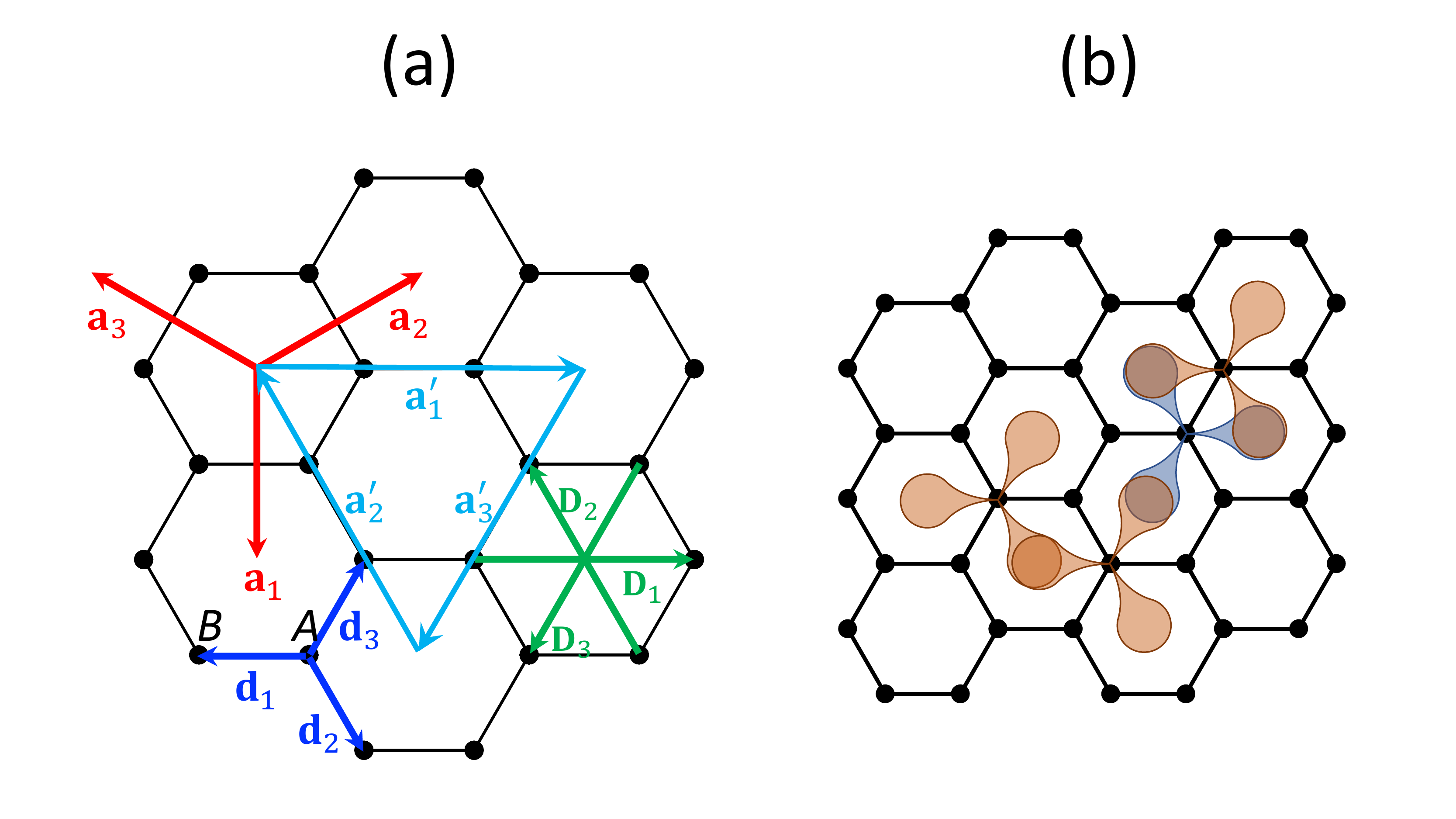}
\end{center}
\caption{(a) Moir\'e superlattice of the TBG: sublattices A and B and 1NN vectors $\vd$ ($%
l=1,2,3 $), 2NN as well as primitive vectors $\va$, and 3NN vectors $\vD$. These vectors are related by $\vd=\frac{1}{3}\left(\vam - \vap\right)$ and $%
\vD=-2\vd$. The 5NN vectors $\va'=\vap - \vam$ will be unit vectors for the Kekul\'e valence bond order. We identify subscripts $l \equiv l~\text{%
mod}~3$ in this paper, for instance $\mathbf{d}_{4}\rightarrow \mathbf{d}_{1}$. (b) Illustration of Wannier orbitals and their overlap with neighbors. A lobe of a Wannier orbital takes charge of $e/3$.}
\label{fig:lattice}
\end{figure}

To deal with the degenerate problem, we adopt
the slave-rotor representation \cite{Florens2002,Florens2004} in which an electron operator $c_{\sigma
}^{\dagger }$ is written in terms of the spinon (auxiliary fermion) $f_{\sigma
}^{\dagger }$ and the rotor $\theta $ conjugate to some charge (or angular
momentum). 
Different from the original setting \cite{Florens2002,Florens2004}, the
charge we concern is the hexagon charge, so we define the angular momentum $%
L_{i_{c}} = Q_{i_{c}} -6$ that equates the hexagon charge relative to six.
Because three hexagons meet at a sublattice site, an electron operator for
any site needs three rotors for adjoining hexagons.
Considering that the lattice is bipartite, we define operators on site A and B as 
\begin{align}
\begin{split}
c_{A \kappa i \sigma }^{\dagger } = f_{A \kappa i \sigma }^{\dagger }
\prod_{l=1,2,3} e^{i \theta_{i- \vd}}, \\
c_{B \kappa i \sigma }^{\dagger } = f_{B \kappa i \sigma }^{\dagger }
\prod_{l=1,2,3} e^{i\theta_{i + \vd}},
\end{split}
\label{creationOP}
\end{align}
where $\mathbf{d}_{1,2,3}$ are three vectors connecting nearest neighbor (NN) A and B sites
[see Fig.~\ref{fig:lattice}(a)], and $e^{\pm i \theta_{i_{c}}}$ will increase
(decrease) the quantum number of $L_{i_{c}}$ by one, i.e. $\left[L_{i_{c}},
e^{\pm i \theta_{i^{\prime }_{c}}}, \right] = \pm e^{\pm i \theta_{i_{c}}}
\delta_{i_{c},i^{\prime }_{c}}$. 
The slave rotor formalism here is designed for the specific filling.
Unlike conventional slave-rotor approach where single fermion Hilbert space on site $i$ is enlarged into
$\mathcal{H}_{\text{f}}(i)\to \mathcal{H}_{\text{spin}}(i)\otimes \mathcal{H}_{\text{rotor}}(i)$. 
The representation is designed such that the cluster energy of the full many-body fermion
wave function is captured by bosonic $U(1)$ rotors, $e^{i\theta_{i_c}}$, 
defined on center of plaquettes. 
Using this representation, we don't have the conventional $U(1)$ gauge
redundancy in contrast with slave-rotor approach for various Hubbard models.
The local gauge symmetry is discussed in Appendix \ref{Appendix:localgauge}.

\begin{figure}[tb]
\begin{center}
\includegraphics[width=0.45\textwidth]{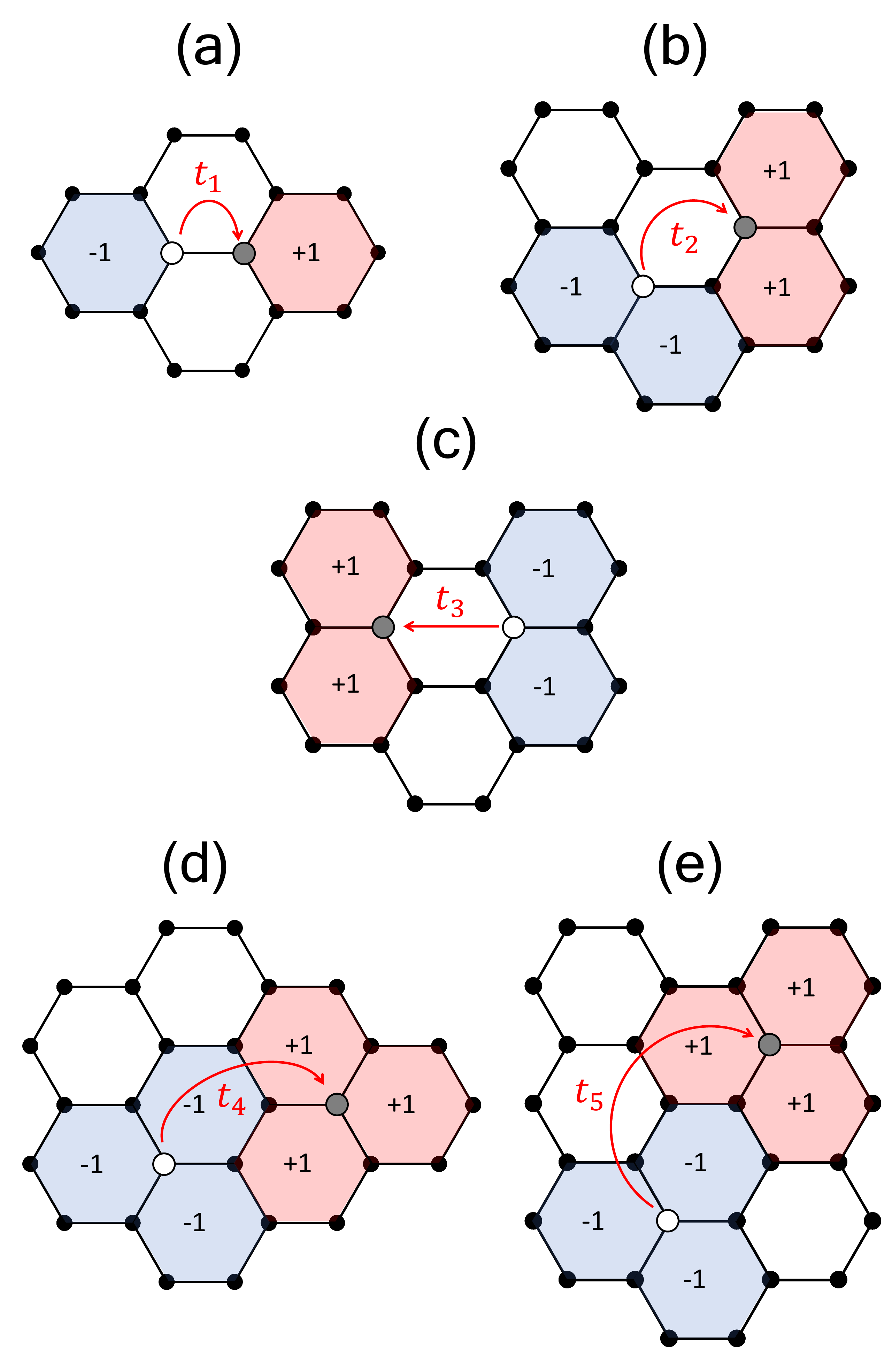}
\end{center}
\caption{Schematic illustration of hexagon charge variation ($\pm1$) due to
electron hopping. (a)-(e) are for 1NN, 2NN, 3NN, 4NN, and 5NN hoppings, respectively. When a hexagon increases (decreases) its charge by one, a
rotor factor $e^{ i \theta}$ ($e^{- i \theta}$) is introduced.
So the hopping renormalization factor in the single-site approximation is $Z=\expval{e^{ i\theta}} 
\expval{e^{- i\theta}}$ in (a), $Z^2$ in (b,c), and $Z^3$ in (d,e).}
\label{fig:phase}
\end{figure}

Taking Eq. (\ref{creationOP}) into $H_{t}$, one observes that there are different
numbers of phase factors in it by referring to Fig.~\ref{fig:phase}.
The NN hopping brings out two phase factors like $e^{i \theta_{i_{c}} -
\theta_{i^{\prime }_{c}}}$ for one hexagon loses a charge and another increases one.
Differently, the 2NN and 3NN hoppings introduce four phase factors.
For rest hoppings of longer distance, six phase factors show up. It is this complicated
$H_t$ in the slave-rotor representation that makes the solving difficult.

The interaction $H_{U} $, on the other hand, is written as $H_{U}=U\sum_{i_{c}} L_{i_{c}} ^{2}
$. In order to impose the constraint on $L_{i_{c}}$, we add 
\begin{equation}
H_h = h \sum_{i_{c}} \left( L_{i_{c}} - \sum_{j \in i_{c}} \sum_{\kappa, \sigma}
f^{\dagger}_{\kappa j \sigma} f_{\kappa j \sigma}+6 \right), \label{constraint_L}
\end{equation} 
to $H$, where $h$ is a Lagrange multiplier determined by optimization, and call the total $H'=H+H_h$.
Considering the particle number constraint, 
\begin{equation} \label{particle_number}
\sum_{\kappa,\sigma} \expval{ f_{\kappa i \sigma}^{\dagger}f_{\kappa i \sigma}} = \sum_{\kappa,\sigma} \expval{ c_{\kappa i \sigma}^{\dagger}c_{\kappa i \sigma} }= 1 + \frac{x}{2},
\end{equation}
Eq. (\ref{constraint_L}) becomes the constraint 
\begin{equation} 
\expval{L_{i_{c}}} = 3x \label{MFconstraint_L}
\end{equation} 
in the mean-field level. Here $x$ is the doping concentration, and $x>0$ ($x<0$) is for electron (hole) doping.

We will solve the model in the slave-rotor representation by using the mean-field theory
to decouple the rotor and the spinon sectors as
$H'=H^{\text{MF}}_{\theta}+H^{\text{MF}}_{f}$, by which the electron ground-state wave
function is the product of those of two sectors
$\ket{\Psi}=\ket{\Psi_{\theta}}\ket{\Psi_{f}}$ subject to the constraints in Eqs.
(\ref{particle_number}) and (\ref{MFconstraint_L}).
The decoupled $H^{\text{MF}}_{\theta}$ \sout{is not solvable} can not be solved
exactly.
We self-consistently solve it by three approximation methods: the single-site, two-site approximation, and the
$O(2)$ nonlinear sigma model and discuss our solutions in the following sections.

\subsection{Single-site approximation} 
\label{sec:single-site}
The single-site approximation is a cluster mean-field theory which approximates that
rotors in different clusters are independent and they interact with the mean-field
environment to gain self-energies~\cite{Florens2004}.
Here the cluster has only one site. 
We emphasize that the "site" is for the rotor as a position of a hexagon when discussing single-site and two-site approximations. 
The mean field turns out to be the rotor's condensate fraction $\sqrt{Z} \equiv \expval{e^{ i\theta_{i_{c}}}} $ and it has no spatial correlation, e.g. $%
\expval{e^{
i\theta_{i_c}} e^{-i\theta_{j_c}}} \approx \expval{e^{
i\theta_{i_c}}}\expval{e^{ -i\theta_{j_c}}}=  Z$. (We choose a real gauge $Z \ge 0$.) Using
$\prod_{i=1}^{m} e^{ i \theta_{i}} e^{ -i \theta_{i'}} \approx m Z^{m-1/2} \left( e^{ i \theta_{i}}+e^{ -i \theta_{i}} \right)$, we have 
\begin{equation}
H^{\text{MF}(1)}_{\theta} = K \left(e^{ i\theta_{i_{c}} } + e^{ -i\theta_{i_{c}} }\right)+ U
L_{i_{c}}^2 + h L_{i_{c}} ,  \label{H1_theta}
\end{equation}
where implicit summation over $i_{c}$ is taken, $K= \left( \sqrt{Z} \sum_j^{\prime} + 2\sqrt{Z^3} \sum_j^{\prime
\prime} + 3\sqrt{Z^5} \sum_j^{\prime \prime \prime} \right) K_{ij} $ from
mean-field of $H_{t}$ with $K_{ij} = \sum_{\kappa} t_{\kappa,ij} \chi^{\kappa}_{ij} + \text{c.c.}$ and $\chi^{\kappa}_{ij} \equiv \sum_{\sigma } 
\expval{ f_{\kappa i \sigma}^{\dagger }f_{\kappa j\sigma }}$ which
renormalization factors depend on distance of $ij$ so we decouple them by $%
\sum_j^{\prime},~\sum_j^{\prime \prime}$, and $\sum_j^{\prime \prime \prime}$
as we elaborated before. 
We replace $\chi^{\kappa}_{ij}$ by $\chi^{\kappa}_{n}$ when $ij$ are nNN sites and the same for $K_{n}$.
When we consider hopping up to 5NN bonds, $K=3\sqrt{Z}M$ with 
\begin{equation}
M=\left(K_1+4ZK_2+2 Z K_3+ 6Z^2 K_4+6Z^2 K_5 \right). \label{M}
\end{equation}

Equation (\ref{H1_theta}) is reduced to a single-site problem, and $Z$ and $h$ have to
be determined self-consistently.
Starting by initializing $K$ and $h$, we write Eq. (\ref{H1_theta}) in the eigenbasis of
$L$, {$\ket{n_{\theta}}$}, in which $e^{ i\theta_{i_{c}}} $ becomes off-diagonal.
Numerically, a truncated Hilbert space of $|n_{\theta}| \le n_{\text{truncate}} $ is
used, which is justified if the results are weakly susceptible to
$n_{\text{truncate}}$.
After diagonalizing and obtaining the ground state of Eq. (\ref{H1_theta}), new
$Z=\expval{e^{ i\theta_{i_{c}}}}^2$ and hence $K$ are generated to update Eq.
(\ref{H1_theta}).
Therefore, it is an iterative procedure to find $K$ or $Z$ until convergence.
Meanwhile, one needs to tune $h$ as well so as to maintain Eq. (\ref{MFconstraint_L}).
As for $\expval{ f_{\kappa i \sigma}^{\dagger }f_{\kappa j\sigma }}$ in $K$, they are
evaluated from the ground state of the spinon sector, which similarly depends on $Z$
learned from the rotor sector.

The mean-field Hamiltonian of the spinon reads
\begin{equation}
H^{\text{MF}(1)}_{f}=\sum_{ ij } \sum_{\kappa,\sigma} \left( t^{\text{eff}%
}_{\kappa, ij} - \mu^{\text{eff}} \delta_{ij} \right)f_{\kappa
i\sigma}^{\dagger }f_{\kappa j\sigma }. \label{H_f}
\end{equation}
Similarly, some parameters in $H^{\text{MF}}_{f}$ are answered by the rotor.
The effective chemical potential is shifted to be $\mu^{\text{eff}}=\mu+3h$, and
the hopping $t^{\text{eff}}_{ij}$ are renormalized by degrees of the rotor,
giving $t^{\text{eff}}_{\kappa, ij} =\left\lbrace Z,Z^2,Z^3\right\rbrace t_{\kappa, ij}
$ for different distances.
Specifically, $t^{\text{eff}}_{\kappa, 1} =Z t_{\kappa, 1} $, $t^{\text{eff}}_{\kappa,
2(3)} =Z^2 t_{\kappa, 2(3)} $, and $t^{\text{eff}}_{\kappa, 4(5)} =Z^3 t_{\kappa, 4(5)}
$.
One solves the ground state of Eq. (\ref{H_f}) with a proper $\mu^{\text{eff}}$ under
the particle-number constraint in Eq. (\ref{particle_number}).
The resulting $\expval{ f_{\kappa i \sigma}^{\dagger }f_{\kappa j\sigma }}$ here are
used to revise those in Eq. (\ref{H1_theta}).
Once again, one returns to the rotor section and solves  until convergent.

\begin{figure}
\begin{center}
\includegraphics[width=0.5\textwidth]{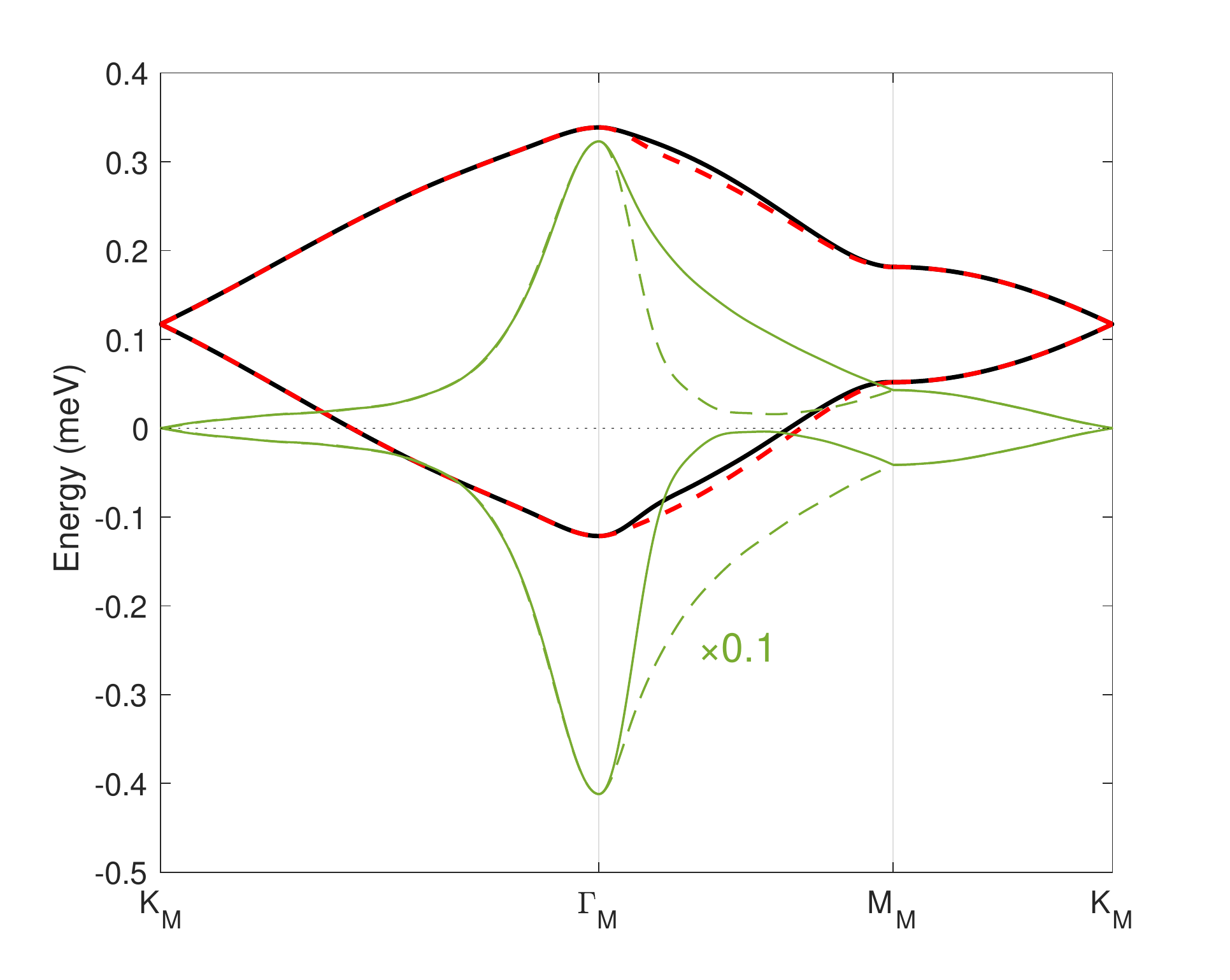}
\caption{Band structures for the renormalization factor $Z=0.205$ (thick black and red lines) and $Z=1$ (thin green lines). The solid and dashed lines represent bands at two valleys $K$ and $K'$, respectively. The renormalized bands are based on the single-site approximation at $U$ and $x=-0.06$. The bare bands ($Z=1$) are scaled by 1/10 to allow for better comparison. The hopping parameters, adopted from Ref. \cite{Koshino2018}, included are up to distance $r<9 a$. }
\label{fig:bands}
\end{center}
\end{figure}

Figure \ref{fig:Z} shows $Z$ (single-site), by the black dotted line, as a function of
$U$ in the undoped case.
The condensate fraction $0 \le Z \le1$ turns out to be the quasiparticle coherent weight.
With increase of $U$, $Z$ decreases and becomes zero when $U$ is larger than the
critical value $U_c \approx 2.8$ meV, above which the system enters the Mott insulator
phase.
However, the observed value of $U_{c}$ is much smaller than the expected value which
should be about the bandwidth of the noninteracting electronic system of 7.35 meV in the
present model.
We also calculated the system with hopping with distance longer than the 5NN site
($r>\sqrt{3} a$), and found an identical $U_{c}$ without prominent change in $Z$ (not
shown).
Figure \ref{fig:bands} shows the dressed bands ($Z=0.205$), which resembles a
graphene-like one, compared to the bare bands ($Z=1$), clearly showing nonuniform band
renormalization as seen in $H_{f}^{\text{MF}(1)}$.
The result justifies it feasible to neglect long-range hopping terms and suggests that
the effective hopping is much shorter with the cluster interaction. 

\subsection{Two-site approximation}
\label{sec:two-site}
In the previous single-site approximation, the quasiparticle weight $Z$ controls
the effective hopping of spinons ($t^{\text{eff}}_{ij}$) and hence the bandwidth
renormalization $Z_{\text{BW}}$ (defined as the ratio of the four-band bandwidth at
finite $U$ to that at zero $U$).
However, it is unphysical that the spinon band becomes completely flat in a Mott
insulator phase because quantum fluctuations can still mediate nonlocal correlations.
Therefore, we have to include spatial corrections among rotors that $\expval{e^{
i\theta_{i_c}} e^{-i\theta_{i'_c}}}\neq 0$ even in the absence of condensation, which
fixes effective hopping amplitudes of the spinon.
This inclusion will solve the problem accordingly, giving finite $Z_{\text{BW}}$ for all
finite $U$. 

We enlarge the cluster to two sites to allow the coupling between two
rotors~\cite{Zhao2007}.
According to Fig. \ref{fig:phase}(a), a NN hopping of an electron creates nonadjacent
rotor excitations, suggesting no correlation between adjacent rotors.
So to decide a two-site cluster, we choose the two sites with one at $i_{c}$ and the
other at $i'_{c}=i_{c}+\mathbf{a}_2^{\prime}$ (we select one of three directions).
To perform the two-site approximation, we single out terms in $H_{t}$ associated with
the target cluster and contract spinon hoppings by $\chi^{\kappa}_{n}$ wherein. 
Then, we do the mean field procedure by contracting spinons as $ \sum_{\sigma } 
f_{\kappa i \sigma}^{\dagger }f_{\kappa j\sigma } \rightarrow \chi^{\kappa}_{ij}$ and
rotors not in the cluster as $e^{i\theta_{j_c}}\rightarrow \sqrt{Z}$.
Consequently, the two-site Hamiltonian of the rotor is 
\begin{align}
\begin{split}
H^{\text{MF}(2)}_{\theta} =&  \frac{1}{2} M \left(e^{ i\theta_{i_{c}} } e^{ -i\theta_{i'_{c}} } + \text{H.c.}\right)  \\ 
& + \frac{5}{2} \sqrt{Z} M \left( e^{ i\theta_{i_{c}} } + e^{ i\theta_{i'_{c}} } + \text{H.c.} \right) \\
&+ U \left( L_{i_{c}}^2 + L_{i'_{c}}^2 \right) + h \left( L_{i_{c}}+L_{i'_{c}}\right),  \label{H2_theta}
\end{split}
\end{align}
where summation over $i_{c}$ is also omitted and $M$ is defined in Eq. (\ref{M}).
(One can check that $H^{\text{MF}(2)}_{\theta}$ becomes $H^{\text{MF}(1)}_{\theta}$ exactly by doing the single-site approximation on it.) 
Numerically, we will solve the ground state in the basis of $\ket{n_\theta,n_{\theta'}}$. 
Because of the coupling of rotors in Eq. (\ref{H2_theta}), we have the spatial
correlation $Z_1 \equiv \expval{e^{ i\theta_{i_{c}} }e^{ i\theta_{i'_{c}}}}$ possibly
different from $Z$.

\begin{figure}
\begin{center}
\includegraphics[width=0.45\textwidth]{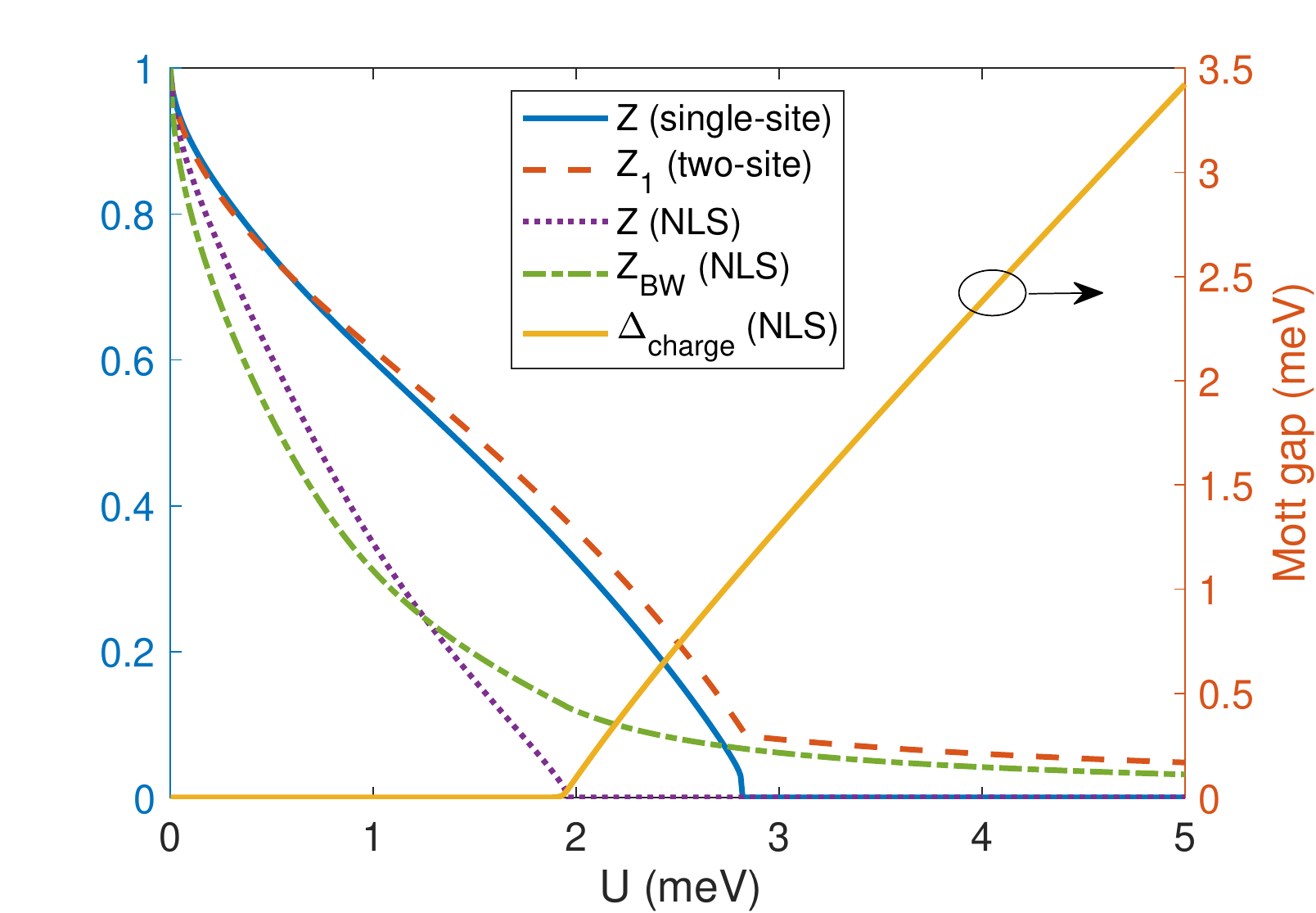}
\caption{Plot of the quasiparticle weight (rotor condensation amplitude) $Z$, the bandwidth renormalization $Z_{\text{BW}}$, and the Mott gap $\Delta_{\text{charge}}$ as a function of $U=5$ meV in the undoped system ($x=0$). Except the dotted line, which is obtained in the single-site approximation, the rest include spatial correlations.}
\label{fig:Z}
\end{center}
\end{figure}

For the spinon Hamiltonian $H^{\text{MF}(2)}_{f}$, it is quite similar to $H^{\text{MF}(2)}_{f}$ but changes the effective hopping integrals as
$t^{\text{eff}}_{\kappa, 1} =Z_1 t_{\kappa, 1} $, $t^{\text{eff}}_{\kappa, 2(3)} =Z_1^2 t_{\kappa, 2(3)} $, and $t^{\text{eff}}_{\kappa, 4(5)} =Z_1^3 t_{\kappa, 4(5)} $.
Similar to what elaborated in Sec. \ref{sec:single-site}, values of $Z$ and $h$ in $H^{\text{MF}(2)}_{\theta}$ and values of $Z_1$ and $\mu^{\text{eff}}$ in $H^{\text{MF}(2)}_{f}$ are solved self-consistently. The result of $Z_1$ is shown in Fig. \ref{fig:Z} by the red dashed line. One can observe that the one-site and two-site approximations have very close $U_{c}$. For the latter, the spatial correlation $Z_1$ at $U<U_{c}$ is contributed mainly from the condensate, and its noncondensate part is maximal at $U_{c}$.

\subsection{Nonlinear sigma model with Hatree-Fock mean field theory}  
\label{sec:NLS}
The last approximation we are going to show is the nonlinear sigma model with minimal
dynamical fluctuations.
We replace $e^{-i\theta_{i_{c}}}$ in $H'$ by the complex bosonic field $X_{i_{c}}$ with
the size constraint $|X_{i_{c}}|^2=1$, which is imposed by a Lagrangian multiplier,
$\lambda$.
This representation transforms the rotor section into an O(2) nonlinear sigma model in
which the angular momentum is the source of dynamical fluctuations of $X$.
The model can be generalized to an O($2N$) model by extending the bosonic field to an
$N$-component complex field $X_{i_{c}, \alpha}$ ($\alpha=1,\ldots,N$) under the
constraint $\sum_{\alpha}|X_{i_{c},\alpha}|^2=N$.
It is know that the $N\rightarrow \infty$ model is solvable for no quantum fluctuations,
so the O(2) model underestimating quantum fluctuations is used to understand the Mott
transition qualitatively.

We construct the Lagrangian in the slave-rotor representation, in which there are
Lagrange multipliers $h$ and $\lambda$ for the constraints.
The angular momentum field $L$ as a conjugate field will be integrated out as we do the replacement $L \rightarrow \partial_{\tau} \theta /U$ in some way. 
To treat the $H_{t}$ term, we repeat the mean field procedure decoupling the spinon and
rotor $H_{t} \rightarrow H_{t,X}^{\text{MF}}+H_{t,f}^{\text{MF}}$, by contracting all
possible correlation functions $\mathcal{Z}_{i_{c}j_{c}} \equiv \expval{X_{i_{c}}^*
X_{j_{c}}}-Z$ and $\chi_{ij}^{\kappa}$ based on the Hatree-Fock principle.
This is based on the assumption that we neglect its dynamical fluctuations. 
Finally, we have the Lagrangians of the rotor and the spinon, and hence Green's functions $G_{X}(i\nu_{n},\vq)$ and $G_{f}^{\kappa}(i \omega_{n},\vk)$.
Owing to their lengthy formulae, we show the details in Appendix \ref{Appendix:NLS}. 

In the mean-field viewpoint, the correlation functions as well as the Lagrange multipliers in the Green's functions take saddle-point values. To have that, these Green's functions must satisfy the equations
\begin{align} 
\expval{|X_{i_{c}}|^2} &= \frac{1}{\beta N_{c}} \sum_{\nu_{n},\vq} G_{X}(i\nu_{n},\vq)= 1, \label{SCE1} \\
\expval{X_{i_{c}}^* X_{j_{c}}} &= \frac{1}{\beta N_{c}} \sum_{\nu_{n},\vq} e^{-i \vq (\vr_{i}-\vr_{j})} G_{X}(i\nu_{n},\vq)= Z+ \mathcal{Z}_{i_{c}j_{c}}, \label{SCE2} \\
\expval{f_{\kappa i \sigma}^{\dagger}f_{\kappa i \sigma} }&= \frac{1}{\beta N_{c}} \sum_{\omega_{n},\vk} e^{i \omega_n 0_{+}}G_{f}^{\kappa}(i \omega_{n},\vk) = \frac{1}{4}\left( 1 - \frac{x}{2} \right), \label{SCE3} \\
\expval{f_{\kappa i \sigma}^{\dagger}f_{\kappa j \sigma} }&= \frac{1}{\beta N_{c}} \sum_{\omega_{n},\vk} 
e^{-i \vk (\vr_{i}-\vr_{j})} G_{f}^{\kappa}(i \omega_{n},\vk) = \chi_{ij}^{\kappa},  \label{SCE4}
\end{align}
which correspond to self-consistent equations for the parameters. 

We show the results in Fig. \ref{fig:Z} by $Z$ (NLS), $Z_{\text{BW}}$ (NLS), and $\Delta_{\text{charge}}$ (NLS).
Both the quasiparticle weight $Z$ and the bandwidth renormalization $Z_{\text{BW}}$
dwindle with $U$ but differently because of the introduction of
$\mathcal{Z}_{i_{c}j_{c}}$ and also the nonlinear band renormalization by
$Z+\mathcal{Z}_{i_{c}j_{c}}$.
In the Mott insulator phase $U>U_c\approx 1.96~\text{meV}$, $Z=0$ while $Z_{\text{BW}}$ remains finite.
The latter, which is close to $Z_{1}$ in the two-site approximation, is quite small; for
example, $Z_{\text{BW}}\approx$ 0.12, 0.06 and 0.03 at $U=2$, 3 and 5 meV,
respectively.
Another important quantity is the excitation gap of the rotor $\Delta_{\text{charge}}$,
attributed to the Mott gap, appearing and growing quite linearly with $U$ in the Mott
insulator phase.
Physically, when the rotor has a gapless spectrum with the minimum at $\vq=\bf{0}$, it
tends to condensate at low temperatures.
Contrarily, a gapful system will forbid the condensation and exhibit short-range
correlations.
Notably, compared present result with previous results, $U_{c}$ is further suppressed.
Now $U_{c}$ is quite closed to $6t_1\approx1.99$ meV, a bandwidth of only 1NN hopping,
inferring that the system is more "flat" than band theory calculations.

\begin{figure}
\begin{center}
\includegraphics[width=0.5\textwidth]{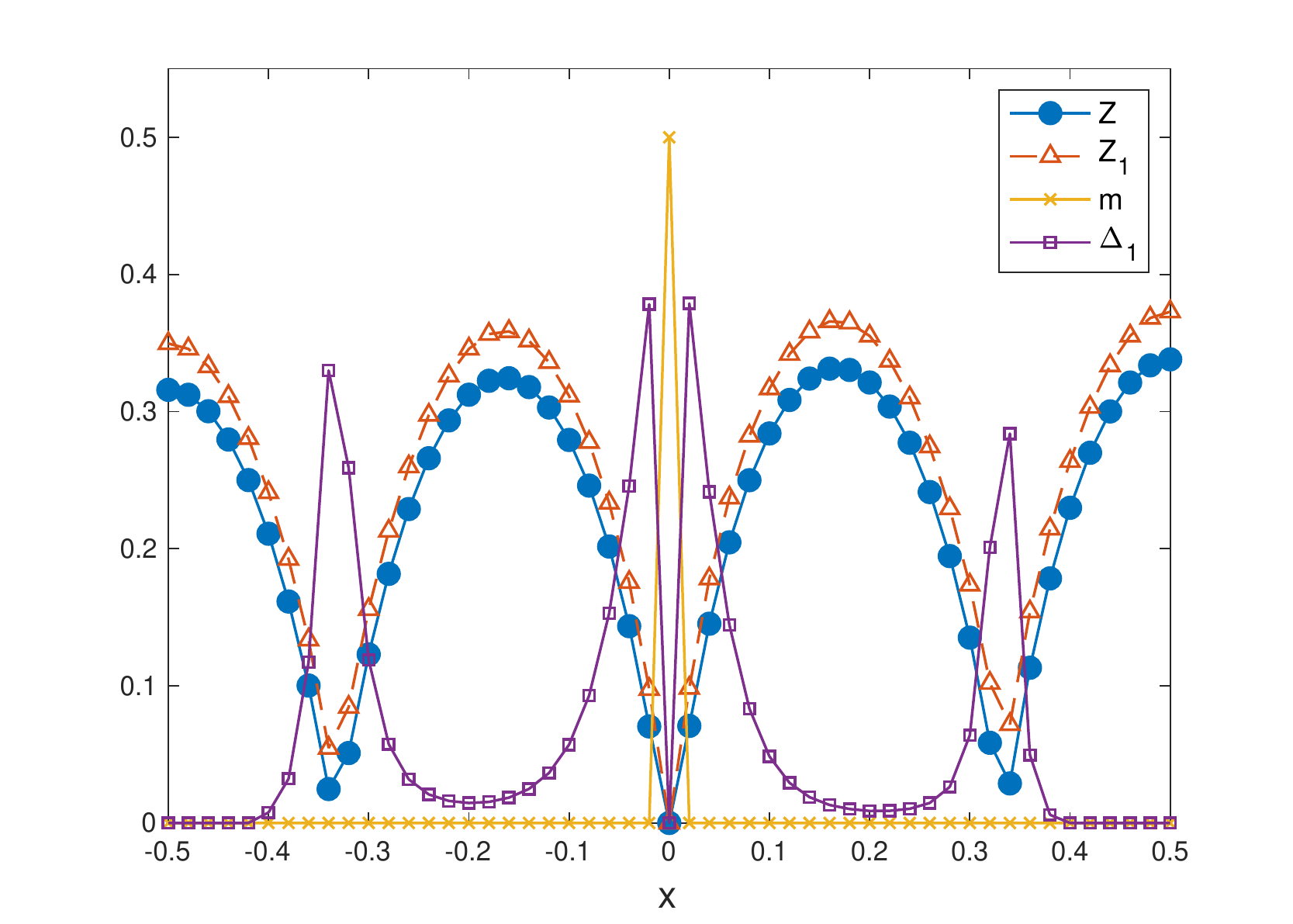}
\caption{Mean-field phase diagram based on the band structure in the two-site approximation at $U=5$ meV and $J_0=0.2$ meV. Mott insulator phases for $Z=0$ appear at $x=0, \pm \frac{1}{3}$.}
\label{fig:phase_diagram}
\end{center}
\end{figure}

\section{Antiferromagnetism and Superconductivity}
\label{sec:AFMSC}
After the study of the Mott physics at $x=0$, we investigate antiferromagnetic (AFM) and
superconducting (SC) instabilities in the cluster Hubbard model.
We will demonstrate that the pair attraction arises naturally through spin/valley
fluctuations in this strongly-correlated and highly-degenerate system.

Our interest is in the strong-correlation regime, so we uncover the instabilities from the large $U/t$ side. 
In the large $U/t$ limit, an effective interaction for the low-energy states, from second order
perturbation, emerges as 
\begin{equation} 
H_{J} = -\frac{1}{2U}\mathcal{P} H_{t} \mathcal{Q} H_{t} \mathcal{P}, 
\end{equation}
where $\mathcal{P}$ is the projection operator to a space of the least
cluster interaction energy, that is $N_{c}(1-|x|)$ hexagons with
six particles and $N_{c}|x|$ hexagons with five particles for
hole doping ($x>0$) or seven particles for electron doping ($x<0$). 
($N_{c}$ is the number of the moir\'e unit cell and $|x|$ is the
doping concentration.) 
Accordingly, $\mathcal{Q}$ projects into a space of
the minimal interaction energy suffered after hopping, which suggests that
the virtual hopping process happens only among hexagons of six charges.
Considering the statistical counting and 1NN hopping contribution
only ($t_1 \in \mathbb{R}$), we approximate 
\begin{equation}
H_{J} \approx -J\sum_{\left\langle ij\right\rangle }
\sum_{\kappa,\kappa^{\prime }} \sum_{\sigma,\sigma^{\prime }} c_{\kappa i
\sigma}^{\dagger }c_{\kappa j \sigma } c_{\kappa^{\prime }j \sigma^{\prime
}}^{\dagger }c_{\kappa^{\prime }i \sigma^{\prime }},  \label{HJ}
\end{equation}%
where $\left\langle ij\right\rangle$ are NN sites and the doping-dependent
coupling is $J=\left(1-|x|\right)^2 J_{0}$.
The coupling constant $J_{0}$ will be regarded as an independent parameter instead of
$t_{1}^2/U$ later.

Since $H_J$ does not create rotor excitations but agitates spinon movement, the spinon operator $f^{\dagger}$ can take over $c^{\dagger}$ there. The mean-field form of $H_{J}$ writes
\begin{equation}
\begin{split}  \label{HJMF1}
H^{\text{MF}}_{J} =& -J \sum_{\vk}\sum_{\kappa,\sigma} \left\lbrace
 \sigma \frac{1}{2}\Delta \sum_{l} e^{i \mathbf{k\cdot d}_l} f^{\dagger}_{\kappa A \vk
\sigma}f^{\dagger}_{\bar{\kappa} B -\vk \bar{\sigma}} + \text{H.c.} \right.
\\
&+ \left. \sigma \kappa \frac{3}{2} m \left( f^{\dagger}_{\kappa A \vk
\sigma}f_{\kappa A \vk \sigma} - f^{\dagger}_{\kappa B \vk
\sigma}f_{\kappa B \vk \sigma} \right) \right\rbrace,
\end{split}
\end{equation}
where two possible orders are introduced: the AFM order $m_{\kappa } = \pm \sum_{\sigma} \sigma \expval{ n_{\kappa i \sigma } } $ and the SC pairing $\Delta^{(l)}=\sum_{\sigma } \sigma 
\expval{
f_{\bar{\kappa} B i+\vd \bar{\sigma }} f_{\kappa A i \sigma }}$.
The AFM order is formed between electrons at the same valley, and $m_{+}=-m_{-}=m$ will
be taken to reflect inversion symmetry.
The SC pairing is formed between electrons at opposite valleys, giving zero net momentum.
The most stable solution we found in the interested region is of $s$-wave (rotationally
invariant), $\Delta^{(l)} = \Delta_s$, spin-singlet and valley-symmetric.
Our solution gives a uniform system without additional periods.
The model does not show the inter-valley coherence wave \cite{Po2018} too because of no
feature of Fermi surface nesting in the band structure \cite{You2019}. 
We remark that we do not extract the Hatree-Fock self-energy in $H_{J}$ to avoid double
counting of interaction since $H_{J}$ originates from $H_{U}$.
The self-energy has been considered in spatial correlations. 

We added $H^{\text{MF}}_{J}$ to $H'$ and solved the model at zero temperature using the
two-site approximation for the rotor section.
The phase diagram at $U=5$ meV and $J_0=0.2$ meV within the doping range $|x|\le 0.5$ is
shown in Fig.~\ref{fig:phase_diagram}.
Repeating dome-shaped $Z$ and $Z_1$ are present and they reach minima at $x= 0,\pm
\frac{1}{3}, \pm \frac{2}{3}, \ldots$, which correspond to the hexagon charge $Q=6,
6\pm1, 6\pm2, \ldots$, respectively.
At these doping concentrations, $Z=0$ for the choice of $U=5$ meV greater than $U_{c}$,
featuring Mott insulation.
The spatial correlation $Z_1$ gives us a sense of the bandwidth about $Z_{1} \times
1.99$ meV.
It would be a defect of the theory that $Z_1$, supposed to be finite, drops to zero at
$x=0$ in the presence of the AFM order. 

The AFM order appears at $x=0$ and is destroyed exceedingly fast with doping.
From the linear gap equation at $T_c$, the critical value of $J$ for the AFM instability
is of order of the Fermi energy, and that explains the quick suppression as
Fig.~\ref{fig:phase_diagram} shows.
As for SC pairing, it appears for the effective attraction respecting the BCS theory and
shows humps centered at Mott insulator phases where the attraction is comparatively
strong.
We found that the spin-singlet valley-symmetric $s$-wave pairing is the most stable
solution in our interesting parameter region.
Since the pairing of spinons in the Mott insulator phases does not indicate
superconductivity because of absence of coherence \cite{Lee2006}, the result infers that
superconductivity is observed noticeably proximity to the Mott insulator phases.

\section{Kekul{\'e} valence bond order}
\label{sec:kekule}
Lastly, we investigate the instability of the lattice.
In this theory, we did not observe a tendency of the intervalley coherence wave or a
$C_3$-symmetry breaking \cite{Po2018,You2019}.
Instead, a KVB is a possible tendency as proposed
by~\citet{xu2018kekule}.
A direct hint of the instability is due to the observation of an unpleasant rotor
dispersion.
In our simulation in Sec. \ref{sec:NLS}, we found that in the Mott phase, there is no
rotor spatial correlations except $\mathcal{Z}_{1} \equiv Z_1-Z$, indicating that rotors
hop on a triangular lattice (more correctly, on three independent lattices) with
distance $\sqrt{3}a = |\mathbf{a}_1^{\prime}|$.
In consequence, the rotor is energetically degenerate at $\Gamma$ and three $K$ points
according to the dispersion $\varepsilon_{X} (\vq) \propto \sum_{l}
\cos(\vq\cdot\va^{\prime})$.
The degeneracy is not stable and should be lift once Bose condensation occurs.
As a result, an order with wave vector
$\mathbf{K}=\frac{2\pi}{3a}(\sqrt{3}\hat{x}+\hat{y})$ might emerge.

Based on this argument, we assume a KVB in the spinon section as 
\begin{equation}
\sum_{\sigma } \expval{ f_{\kappa A i \sigma}^{\dagger }f_{\kappa B i+\vd \sigma }} 
= \chi_1 + \chi_1^{\prime} \cos\left[ \mathbf{K}\cdot (\vr_i-\vd)\right].
\end{equation}
The order is shown in Fig. \ref{fig:phase_diagram} where thick and thin lines indicate
strong bonds $\chi_{+}=\chi_1+\chi_{1}^{\prime}$ and weak bonds
$\chi_{-}=\chi_1-\frac{1}{2}\chi_{1}^{\prime}$, respectively.
The KVB preserves the $C_3$ symmetry and has a threefold enlarged unit cell containing
three hexagons labelled by 1, 2, and 3 in Fig.~\ref{fig:phase_diagram}.
The modulation in the spinon section will induce the order in the rotor section as
well.
We define the rotor's condensate fraction $\expval{e^{i \theta_{i_c}}}$ as
$\sqrt{Z_{+}}$ at site $1$ and as $\sqrt{Z_{-}}$ at site 2 and 3, respectively.
Similarly, the rotor correlation $\expval{e^{i \theta_{i_c}} e^{-i \theta_{j_c}} }$ as
$Z_{1+}$ for $(i_c,j_c)=(1,1')$ and $Z_{1-}$ for $(i_c,j_c)=(2,2')$ and $(3,3')$.

With this assumption, we repeat the self-consistent calculation using the two-site
approximation.
Due to computational limit, in this section, we neglect long-range hopping except the 1NN
hopping ($t_1$), or a much larger cluster for the correlation is needed.
Meanwhile, on top of the KVB, we investigate AFM and SC instabilities from $H_{J}$ in
Eq. (\ref{HJ}).
The AFM order is defined as before in Sec. \ref{sec:AFMSC}, while the SC order is 
$\sum_{\sigma } \sigma \expval{
f_{\bar{\kappa} B i+\vd \bar{\sigma }} f_{\kappa A i \sigma }} = \Delta_s + \Delta_s^{\prime} \cos\left[ \mathbf{K}\cdot (\vr_i-\vd)\right]$. (The $s$ wave is still stronger than the $d+id$ wave.) 

\begin{figure}
\begin{center}
\includegraphics[width=0.5\textwidth]{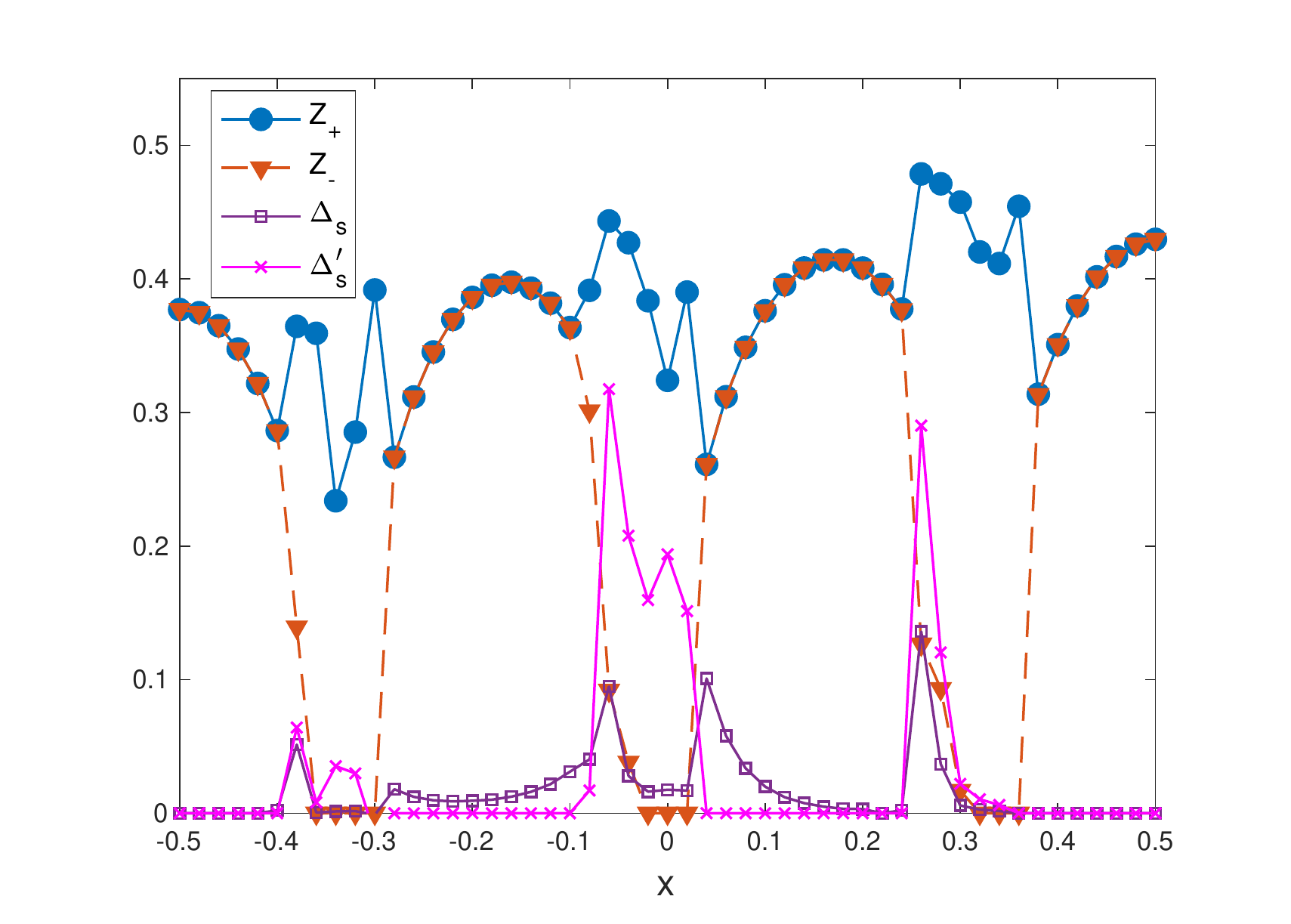}
\caption{Phase diagram of the Kekul{\'e} valence bond order as well as $s$-wave superconductivity at $U=5$ meV and $J_0=0.2$ meV when only the 1NN hopping is considered. When $Z_{+} \neq Z_{-}$ or $\Delta_{s}^{\prime} \neq 0$, the Kekul{\'e} valence bond order is present.}
\label{fig:KVB_phase}
\end{center}
\end{figure}

The phase diagram as doping at $U=5$ meV and $J_0=0.2$ meV is shown in
Fig.~\ref{fig:KVB_phase}, in which we find the KVB occurs around the fractional doping
$x=0,~\pm \frac{1}{3}$, where $Z_{+}\neq Z_{-}$.
The SC order shows strong modulation $\Delta_{s}^{\prime}>\Delta_{s}$ on top of the KVB
order, and the AFM order is unfavorable completely.
Remarkably, the condensate fraction $Z_{-}$ is zero for a finite doping range nearby the
fractional doping, whereas $Z_{+}$ is finite.
At these doping concentrations, the system is an insulator phase, named a Kekul{\'e}
valence bond solid, because charge is localized at hexagons 2 and 3 and cannot
propagate.
As a result, the KVB will extend the Mott insulator phase, giving us a phase diagram
quite similar to what experiments
observed~\cite{cao2018unconventional,cao2018correlated}.
Whether the KVB is the origin of the correlated phase awaits future investigation.

\begin{figure}
\begin{center}
\includegraphics[width=0.2\textwidth]{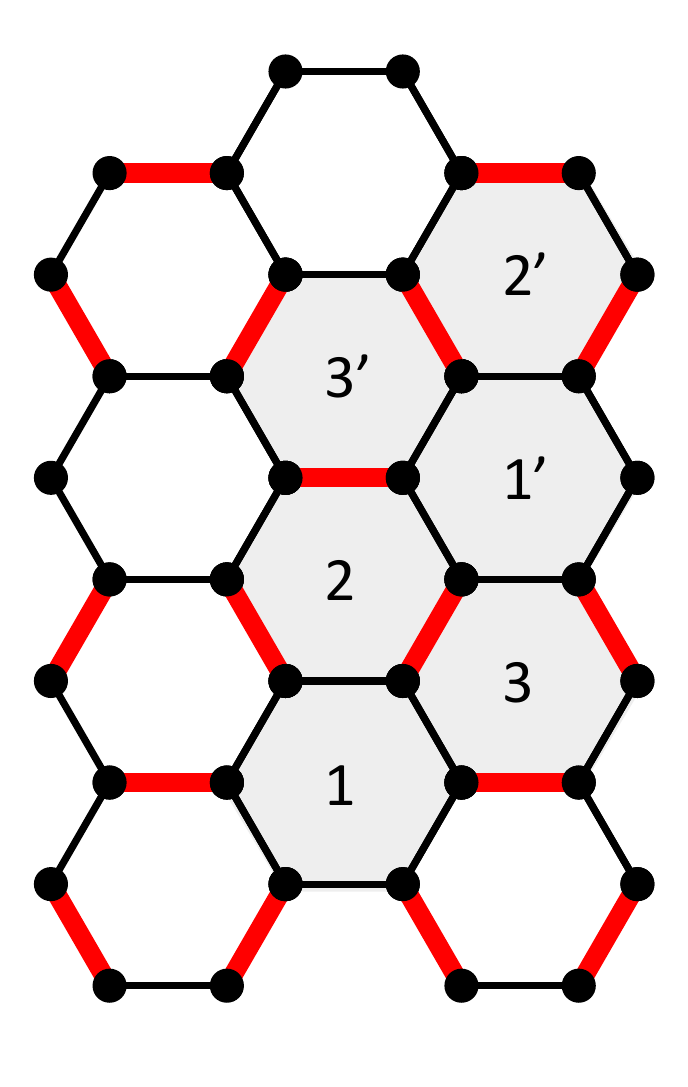}
\caption{Kekul{\'e} valence bond order as a modulation of hopping amplitudes with strong NN bonds (red thick lines) and weak NN bonds (black thin lines). The order enlarges the unit cell containing three hexagons labelled by 1, 2, and 3. Three pairs of hexagons [(1,1'), (2,2'), and (3,3')] are considered to study rotor correlation.}
\label{fig:phase_diagram_Kekule}
\end{center}
\end{figure}

\section{Conclusions}
\label{sec:conclusion}
In this work, we have applied the slave-rotor theory on the cluster Hubbard model based
on the peculiar Wannier orbitals of magic-angle TBG.
The slave-rotor will play the role of the charge degree of freedom in a hexagon (six
sites).
Due to strong correlation, charge fluctuation is suppressed to exhibit Mott physics.
More than what
suggested~\cite{Xie2018nature,Ochi2018,choi2019electronic,polshyn2019large}, the theory
predicts multiple Mott insulator phases at fractional filling with $n=2$ (two charges
per moir\'e unit cell), $2\pm \frac{1}{3}$, $2\pm \frac{2}{3}$, $2\pm 1$ etc.
In addition, the theory suggests a Kekul{\'e} valence bond solid phase~\citet{xu2018kekule} nearby the fractional filling.
Experimental results from scanning tunnelling microscopy showed prominent depression of
spectrum at $n=2$ and weaker ones $n=1$ and $3$
\cite{cao2018unconventional,Xie2019,choi2019electronic,polshyn2019large,Lu2019};
However, exotic dips seemed present in between, which awaits attentive examination;
similar features were seen in transport measurements~\cite{Lu2019}.
With doping away from the Mott insulator, due to spin-valley fluctuations $s$-wave
superconductivity emerges naturally by breaking approximate SO(4)
symmetry~\cite{You2019}.
Lastly, we comment that our slave-rotor theory for the system does not enumerate cluster
states which have discriminative roles in hopping~\cite{Bunemann1998}.
Many nontrivial phases~\cite{Sharpe2019,Lu2019} might be due to the nonlocal cluster
interaction, and advanced techniques for cluster theories will be
required~\cite{Maier2005,Medici2009,Lee2017}.
Nevertheless, our theory incorporating Mott physics provides an important ground for the
highly correlated and degenerate system. 

\begin{acknowledgments}
S.M.H and T.K.L thank F. Yang for helpful discussions. S.M.H. is supported by the
Ministry of Science and Technology (MoST) in Taiwan under Grant No.105-2112-M-110-014-MY3 and No. 108-2112-M-110-013-MY3. Y.-P.H receives funding from the European Union's Horizon 2020 research and innovation program under the Marie Sk\l odowska-Curie grant agreement No. 701647.
\end{acknowledgments}

\appendix
\section{Local gauge symmetry}  
\label{Appendix:localgauge}
We demonstrate the local gauge symmetry present in this slave-rotor theory.
In a typical slave-particle theory, taking the slave-boson theory for
example~\cite{Lee2006}, an electron operator $c$ is written as a product of a boson $b$
and a fermion operator $f$: $c_{i}^{\dagger}=b_{i} f_{i}^{\dagger}$.
Because both $f$ and $b$ locate at the same site $i$, one can understand a local gauge symmetry in the theory that under the local gauge transformation
\begin{equation}
f_{i}^{\dagger} \rightarrow e^{i\phi_{i}} f_{i}^{\dagger} , \quad b_{i} \rightarrow e^{-i\phi_{i}}  b_{i} \notag
\end{equation} 
the system is invariant. 
The local gauge symmetry implies particle conservation that the net current of $f$ and
$b$ particles is zero: ${j}_{f}^{\mu}+{j}_{b}^{\mu}=0$ and an internal gauge field
$a_{\mu}$ couples to $f$ and $b$ particles simultaneously ($\mu$ for spacetime
labels)~\cite{Lee2006}.

In our slave-rotor theory, a rotor stands for the charge in a hexagon which shares with
many spinons at the corner, so that it is impossible to find a gauge-invariant
transformation just on a rotor and a spinon.
Differently, we will show the local gauge transformation is on the spinon and rotor
dipoles.
The existence of a local gauge symmetry is because we can find that a hop of an electron
as well as a spinon always accompanies with a hop of a rotor (hexagon charge), an
indication of charge conservation.
Besides, according to the definition in Eq. (\ref{creationOP}), an electron creation
operator takes three rotor raising operators, so a phase operator $e^{i \theta}$ create
a charge of 1/3.

Let's discuss the transformations one by one.
Firstly, refer to Fig. \ref{fig:phase}(a).
It describes a NN hopping as $c_{\vr_{i}}^{\dagger}c_{\vr_{i}+ \vd}$ (with $l=1$), which
becomes $f_{\vr_{i}}^{\dagger}f_{\vr_{i}+ \vd} X_{\vr_{i}- \vd}^{*} X_{\vr_{i}+ 2\vd}$
after substitution, where we omit the spin-valley index and use $X_{i_{c}}^{*} \equiv
e^{i\theta_{i_{c}}}$.
The four-operator term possesses a local gauge symmetry under the transformation:
\begin{align}
f_{\vr_{i}}^{\dagger}f_{\vr_{i}+ \vd} &\rightarrow e^{i\phi_{\vr_{i}+ \frac{1}{2}\vd}} f_{\vr_{i}}^{\dagger}f_{\vr_{i}+ \vd} , \label{spinondipole1} \\
X_{\vr_{i}- \vd}^{*} X_{\vr_{i}+ 2\vd} &\rightarrow e^{-i\phi_{\vr_{i}+ \frac{1}{2}\vd}}  X_{\vr_{i}- \vd}^{*} X_{\vr_{i}+ 2\vd}, \label{rotordipole}
\end{align}
where the gauge phase is define at the center of the NN bond $\vr_{i}+ \frac{1}{2}\vd$.
We can interpret the above equations as a gauge transformation on a spinon and a rotor
dipole.
The rotor dipole and its transformation will be an elementary one as seen later.
As the rotor dipole is threefold the length of that of the spinon dipole, the rotor's
velocity is three times as big as the spinon's in this hopping, agreeing with equal
currents of the spinon and the rotor. 

Next, we consider the gauge transformation in Fig. \ref{fig:phase}(b), where we can find
a spinon dipole of length $a$ and two rotor dipoles of length $3|\vd|$.
To make consistent with before, if the rotor dipoles transform according to Eq. (\ref{rotordipole}), the spinon dipole of length $a$ has to transform as
\begin{equation}
f_{\vr_{i}}^{\dagger}f_{\vr_{i}- \va} \rightarrow e^{i\phi_{\vr_{i}- \frac{1}{2}\vdp}} e^{i\phi_{\vr_{i}-\vdp+ \frac{1}{2}\vdm}} f_{\vr_{i}}^{\dagger}f_{\vr_{i}- \va} , \label{spinondipole2}
\end{equation}
(with $l=2$) where the gauge phase connecting $\vr_{i}$ and $\vr_{i}- \va$ is defined as
an addition of a gauge phase at two bond centers $\vr_{i}- \frac{1}{2}\vap$ and $\vap+
\frac{1}{2}\vam$. 

Lastly, we discuss the 3NN hopping case in Fig. \ref{fig:phase}(c), where
$c_{\vr_{i}}^{\dagger}c_{\vr_{i}+ \vD}$ ($l=1$) arises.
In Fig. \ref{fig:phase}(c), there are two rotor dipoles transformed according to Eq.
(\ref{rotordipole}).
To make gauge invariant, the spinon dipole of length $|\vD|$ should follow the transformation:
\begin{equation}
f_{\vr_{i}}^{\dagger}f_{\vr_{i}+ \vD} \rightarrow e^{i\phi_{\vr_{i}+\vdp- \frac{1}{2}\vd}} e^{i\phi_{\vr_{i}+\vdm- \frac{1}{2}\vd}}  f_{\vr_{i}}^{\dagger}f_{\vr_{i}+ \vD} , \label{spinondipole3}
\end{equation}
where the two gauge phases are conjugates of those of associated rotor dipoles.
As the gauge transformations in the rest hopping terms, readers can use and combine the
transformations in Eqs. (\ref{spinondipole1})-(\ref{spinondipole3}).

\section{Slave-rotor representation for $H_{t}$} 
\label{Appendix:MF}
In this Appendix, we are aim at the hopping Hamiltonian,
\begin{equation}
H_{t}=\sum_{ij}\sum_{\kappa =\pm} \sum_{\sigma =\uparrow ,\downarrow
} t^{\kappa}_{ij} c_{\kappa i \sigma }^{\dagger
}c_{\kappa j \sigma }, \label{Ht_appendix}
\end{equation} 
and transform physical electrons ($c^{\dagger}$) into the spinon ($f^{\dagger}$) and the
slave rotor ($e^{i \theta}$) operators.
We demonstrate them up to the fifth NN one. 
For brevity, we denote an n-th NN hopping by nNN and their corresponding hopping
integrals are $t_{n}$.
The honeycomb lattice shows that two sublattices A and B have different hexagon
neighbors, so we define that $c_{\kappa A i \sigma }^{\dagger } = f_{\kappa A i \sigma
}^{\dagger } \prod_{l} e^{i \theta_{i- \vd}} $ and $c_{\kappa B i
\sigma }^{\dagger } = f_{\kappa B i \sigma }^{\dagger } \prod_{l} e^{i \theta_{i + 
\vd}} $. 
In this formalism, a 1NN brings out two phase factors, while 2NN and 3NN give four.
As for 4NN and 5NN as well as longer-ranged hoppings, they give six phase factors.

In addition, we will decompose the spinon and the rotor by the Hatree-Fock mean field
theory to have a bilinear form of the hopping Hamiltonian.
The complex bosonic field $X\equiv e^{-i\theta}$ will be used.
For the simplest mean-field, one is to take $\expval{X}=\sqrt{Z}$ for condensation of
the rotor.
Here we are going to consider the spatial correlations from excitations of the rotors.
In the spirit of Bose condensation, we write $X=\sqrt{Z}+\delta X$, the latter for the
non-condensate component.
We define the correlation functions for the spinon and the rotor as
\begin{align}
\chi_{\text{r}}^{\kappa} &= \sum_{\sigma} \expval{ f_{\kappa i\sigma }^{\dagger }f_{\kappa i+\vr\sigma } }, \\
\mathcal{Z}_{\vR}&=\expval{\delta X_{i_{c}}^{*} \delta X_{i_{c}+\vR} }=\expval{\delta X_{i_{c}} \delta X_{i_{c}+\vR}^{*} }.
\end{align} 
Since rotors are assigned to hexagon sites, $\vR$ in $\mathcal{Z}_{\vR}$ has to be a
crystal translation vector, while $\vr$ in $\chi_{\vr}$ depends on whether it is an
inter or intra-sublattice bond.
Because of spatial symmetry, we further denote some correlation functions by
\begin{align}
\begin{split}
&\chi^{\kappa}_{1} = \chi^{\kappa}_{\vd}, \:\chi^{\kappa}_{2} = \chi^{\kappa}_{\pm \va}, \:\chi^{\kappa}_{3} = \chi^{\kappa}_{ \vD}, \\ 
&\chi^{\kappa}_{4} = \chi^{\kappa}_{\vd-\vap}= \chi^{\kappa}_{\vd+\vam}, \\
&\chi^{\kappa}_{5} = \chi^{\kappa}_{\pm \left( \va- \vap \right)}, \\
& \mathcal{Z}_0 = \mathcal{Z}_{\pm \va}, \: 
 \mathcal{Z}_1 = \mathcal{Z}_{\pm \left( \va-\vap \right)}, \: 
 \mathcal{Z}_2 = \mathcal{Z}_{\pm 2\va}, \\
& \mathcal{Z}_3 = \mathcal{Z}_{\pm \left( 2\va-\vap \right)}
= \mathcal{Z}_{\pm \left(\va-2\vap \right)},
\end{split}
\end{align}
where $l=1,~2,~3$.

After tedious derivation of the mean-field contraction and also Fourier transform, we have the decoupled mean-field Hamiltonians of the spinon and the rotor $H^{\text{MF}}_{t}= H^{\text{MF}}_{t,f}+H^{\text{MF}}_{t,X}$ to bilinear order:
\begin{equation}
\begin{split}
H^{\text{MF}}_{t,f}=& \sum_{\vk} \sum_{\kappa,\sigma} \left\lbrace h_{0,\kappa}(\vk) 
\left( f_{\kappa A \vk \sigma}^{\dagger}f_{\kappa A \vk \sigma} +f_{\kappa B \vk \sigma}^{\dagger}f_{\kappa B \vk \sigma} \right) \right. \\
& + \left. 
h_{1,\kappa}(\vk)  f_{\kappa A \vk \sigma}^{\dagger}f_{\kappa B \vk \sigma} 
+ h^{*}_{1,\kappa}(\vk)  f_{\kappa B \vk \sigma}^{\dagger}f_{\kappa A \vk \sigma}
\right\rbrace,
\end{split}
\end{equation}
and
\begin{equation}
H^{\text{MF}}_{t,X}= \sum_{\vq} \varepsilon_{X} (\vq) X_{\vq}^{*} X_{\vq},
\end{equation}
where
\begin{widetext}
\begin{equation} \label{h0}
\begin{split}
h_{0,\kappa}(\vk)  & = 2  t_2^{\kappa}\left[ Z^2+Z\left(\mathcal{Z}_0+2\mathcal{Z}_1+\mathcal{Z}_2\right) + \mathcal{Z}_1^2+\mathcal{Z}_0 \mathcal{Z}_2\right] \sum_{l}  \cos(\vk\cdot\va) \\
&\quad + 2 t_5^{\kappa}\left[ Z^3+Z^2 \left( 2\mathcal{Z}_0 + 3\mathcal{Z}_1 + 2\mathcal{Z}_2 + 2\mathcal{Z}_3 \right) \right. 
\\
&\quad \quad +Z\left( 3\mathcal{Z}_1^2+\mathcal{Z}_2^2 + 2\mathcal{Z}_0 \mathcal{Z}_1 +2 \mathcal{Z}_0 \mathcal{Z}_2 + 3\mathcal{Z}_0 \mathcal{Z}_3 +2\mathcal{Z}_1\mathcal{Z}_2 + 2\mathcal{Z}_1 \mathcal{Z}_3 +2 \mathcal{Z}_2 \mathcal{Z}_3 \right) \\
 &\quad \quad + \left. \mathcal{Z}_1^3+\mathcal{Z}_1 \mathcal{Z}_2^2+ 2\mathcal{Z}_0 \mathcal{Z}_1 \mathcal{Z}_3 + 2\mathcal{Z}_0 \mathcal{Z}_2\mathcal{Z}_3 \right]  \sum_{l} \cos\left( \vk\cdot\va^{\prime} \right) \\
 &= 2  t_2^{\kappa}\left(Z+\mathcal{Z}_1\right)^2 \sum_{l}  \cos(\vk\cdot\va) 
 + 2 t_5^{\kappa}\left( Z+\mathcal{Z}_1\right)^3 \sum_{l} \cos\left( \vk\cdot\va^{\prime} \right)
\end{split}
\end{equation}
\begin{equation} \label{h1}
\begin{split}
h_{1,\kappa}(\vk)  & =  t_1^{\kappa} (Z+\mathcal{Z}_1) \sum_{l}  e^{i \vk\cdot\vd}
+ t_3^{\kappa}\left[ Z^2+Z\left(2\mathcal{Z}_1+2\mathcal{Z}_2 \right) + \mathcal{Z}_1^2+\mathcal{Z}_2^2 \right] \sum_{l}  e^{i \vk\cdot\vD}  \\
& \quad + t_4^{\kappa}\left[Z^3+Z^2 \left( 3\mathcal{Z}_0 + 3\mathcal{Z}_1+2\mathcal{Z}_2+\mathcal{Z}_3 \right) \right. \\
& \quad \quad + Z\left( \mathcal{Z}_0^2+3\mathcal{Z}_1^2+\mathcal{Z}_2^2+3\mathcal{Z}_0\mathcal{Z}_1+4\mathcal{Z}_0\mathcal{Z}_2+3\mathcal{Z}_0 \mathcal{Z}_3 + 2\mathcal{Z}_1\mathcal{Z}_2+\mathcal{Z}_1\mathcal{Z}_3 \right) \\
& \quad \quad + \left. \mathcal{Z}_1^3+\mathcal{Z}_0 \mathcal{Z}_2^2+ \mathcal{Z}_0 ^2 \mathcal{Z}_3 + 2 \mathcal{Z}_1 \mathcal{Z}_0 \mathcal{Z}_2 + \mathcal{Z}_1 \mathcal{Z}_0 \mathcal{Z}_3 \right]
 \sum_{l} e^{i \vk\cdot\vd} \left( e^{-i\vk\cdot\vap}+ e^{i \vk\cdot\vam} \right) \\
& = t_1^{\kappa} (Z+\mathcal{Z}_1) \sum_{l}  e^{i \vk\cdot\vd} 
+ t_3^{\kappa}\left( Z + \mathcal{Z}_1 \right)^2 \sum_{l}  e^{i \vk\cdot\vD}   + t_4^{\kappa}\left(Z + \mathcal{Z}_1 \right)^3 
 \sum_{l} e^{i \vk\cdot\vd} \left( e^{-i\vk\cdot\vap}+ e^{i \vk\cdot\vam} \right)
\end{split}
\end{equation}
and 
\begin{equation} \label{eX}
\begin{split}
\varepsilon_{X} (\vq) 
& = \left\lbrace 2(Z+\mathcal{Z}_2) K_2+2 \left[ 3Z^2+Z(2\mathcal{Z}_0+3\mathcal{Z}_1+3\mathcal{Z}_2+3\mathcal{Z}_2)+2\mathcal{Z}_1 \mathcal{Z}_2 + \mathcal{Z}_1 \mathcal{Z}_3 + 2\mathcal{Z}_0 \mathcal{Z}_3 +\mathcal{Z}_2^2 \right] K_4  \right. \\
& \quad \quad + \left. 4\left[ Z^2+Z(\mathcal{Z}_1+\mathcal{Z}_2+2\mathcal{Z}_3)+\mathcal{Z}_1 \mathcal{Z}_3 + \mathcal{Z}_2 \mathcal{Z}_3 \right] K_5 
\right\rbrace
 \sum_{l}  \cos(\vq\cdot\va) \\
& \quad + \left\lbrace K_1 + 4(Z+\mathcal{Z}_1)K_2 +2(Z+\mathcal{Z}_1) K_3 \right. \\
& \quad \quad+ 2\left[ 3Z^2+Z(3\mathcal{Z}_0+6\mathcal{Z}_1+2\mathcal{Z}_2+\mathcal{Z}_3)+3\mathcal{Z}_1^2+2\mathcal{Z}_0\mathcal{Z}_2+\mathcal{Z}_1\mathcal{Z}_3 \right] K_4 \\
& \quad \quad + \left. 2\left[ 3Z^2+Z(2\mathcal{Z}_0+6\mathcal{Z}_1+2\mathcal{Z}_2+2\mathcal{Z}_3) + 3\mathcal{Z}_1^2+\mathcal{Z}_2^2+2\mathcal{Z}_0\mathcal{Z}_3\right] K_5  
\right\rbrace
\sum_{l}  \cos\left( \vq \cdot\va^{\prime} \right)  \\
&\quad  +  \left\lbrace 2(Z+\mathcal{Z}_0) K_2 + 2(Z+\mathcal{Z}_2) K_3 \right. \\
& \quad \quad + 4\left[ Z^2+Z(2\mathcal{Z}_0+\mathcal{Z}_1+\mathcal{Z}_2)+\mathcal{Z}_0 \mathcal{Z}_2+\mathcal{Z}_1\mathcal{Z}_0 \right] K_4 \\
& \quad \quad + \left. 4\left[ Z^2+ Z(\mathcal{Z}_0+\mathcal{Z}_1+\mathcal{Z}_2+\mathcal{Z}_3) + \mathcal{Z}_1\mathcal{Z}_2+\mathcal{Z}_0 \mathcal{Z}_3 \right] K_5  \right\rbrace 
\sum_{l}  \cos(\vq\cdot 2\va)  \\
& \quad + \left\lbrace \left[ Z^2+Z(3\mathcal{Z}_0+\mathcal{Z}_1)+\mathcal{Z}_0^2 +\mathcal{Z}_0\mathcal{Z}_1 \right] K_4 \right. \\
& \quad  \quad+ \left. 2\left[ Z^2+Z(2\mathcal{Z}_0+\mathcal{Z}_1+\mathcal{Z}_2)+\mathcal{Z}_0 \mathcal{Z}_1+\mathcal{Z}_0 \mathcal{Z}_2 \right] K_5 \right\rbrace
\sum_{l}  \left( \cos \left[ \vq\cdot(2\va-\vap) \right]+\cos \left[ \vq\cdot(\va-2\vap) \right] \right) \\
& = 2Z\left[K_2 + (Z+\mathcal{Z}_1)(3K_4+2K_5)\right] \sum_{l} \cos(\vq\cdot\va) \\
& \quad+ \left[K_1+(Z+\mathcal{Z}_1)(4K_2+2K_3)+6(Z+\mathcal{Z}_1)^2(K_4+K_5) \right] \sum_{l} \cos(\vq\cdot\va^{\prime}) \\
& \quad+2Z\left[ K_2+K_3+2(Z+\mathcal{Z}_1)(K_4+K_5)\right] \sum_{l} \cos(\vq\cdot 2\va) \\
& \quad+ Z(Z+\mathcal{Z}_1)(K_4+2K_5) \sum_{l}  \left( \cos \left[ \vq\cdot(2\va-\vap) \right]+\cos \left[ \vq\cdot(\va-2\vap) \right] \right)
\end{split}
\end{equation}
\end{widetext}
with 
\begin{equation}
K_{i=1,2,3,4,5} = \sum_{\kappa} t^{\kappa}_{i} \chi^{\kappa}_{i} + \text{c.c.} = 4 \Re\left( t^{\kappa}_{i} \chi^{\kappa}_{i}\right).
\end{equation}
The last lines of Eqs. (\ref{h0}-\ref{eX}) show the results when
$\mathcal{Z}_0=\mathcal{Z}_2=\mathcal{Z}_3=0$.
The above formulae, for brevity, omit the rotor's anomalous correlation $\expval{\delta
X_{i_{c}} \delta X_{i_{c}+\va} }$, which might emerge when $Z\neq 0$, for its make the
formulae much lengthy but its smallness is actually ineffective.

We note that our simulation results showed $\mathcal{Z}_0=\mathcal{Z}_2=\mathcal{Z}_3=0$
and only $\mathcal{Z}_1$ being finite in the Mott insulator phase ($Z=0$).
The result is reasonable because the former order parameters do not appear linearly in
$H^{\text{MF}}_{t}$ and $H^{\text{MF}}_{X}$; solutions of this type system are zeros
commonly and become finite through a discontinuous first-order phase transition when
some couplings are greater than critical values.

\section{Action in slave-rotor representation} 
\label{Appendix:NLS}
The action of the system in terms of the rotor and the spinon particles will be
demonstrated.
The Hatree-Fock mean field theory will be adopted to deal with the hopping Hamiltonian
$H_{t}$.
The adoption of the mean field theory leads to several self-consistent equations, which
will be shown here as well.
Since the presentation here is aimed at the Mott transition, it will not include
antiferromagnetism and superconductivity.

The imaginary-time action reads
\begin{equation}
\begin{split}
S&=\int_{0}^{\beta} d\tau  \left\lbrace -i \sum_{i_{c}}  L_{i_{c}}\partial_{\tau}\theta_{i_{c}} 
+ \sum_{i,\kappa, \sigma} \bar{f}_{\kappa i \sigma} \partial_{\tau} f_{\kappa i \sigma} \right. \\
&+ H_{t} + U \sum_{i_{c}} L_{i_{c}}^2 +  \left. h \sum_{i_{c}} \left( L_{i_{c}} - \sum_{i \in i_{c}} \bar{f}_{\kappa i \sigma} f_{\kappa i \sigma} + 6 \right) \right\rbrace,
\end{split}
\end{equation}
where $i$ and $i_{c}$ run over sublattice and hexagon sites, respectively.
The hopping Hamiltonian $H_t$, not shown explicitly here, is referred to Appendix
\ref{Appendix:MF}. 
The conjugate field angular momentum $L_{i_{c}}$ will be integrated out to change to $\partial_{\tau}\theta_{i_{c}}$, giving 
\begin{equation}
\begin{split}
S&=\int_{0}^{\beta} d\tau  \left\lbrace \sum_{i,\kappa, \sigma} \bar{f}_{\kappa i \sigma} \left( \partial_{\tau} -3 h\right) f_{\kappa i \sigma} + H_{t} \right. \\
& \left. + \frac{1}{4U} \sum_{i_{c}} \left(\partial_{\tau} \theta_{i_{c}}+ih\right)^2 + 6h N_{c}  \right\rbrace.
\end{split}
\end{equation}
Now, we will replace $e^{-i\theta_{i_{c}}}$ by $X_{i_{c}}$ with the constraint
$|X_{i_{c}}|^2=1$, which is realized with the aid of the Lagrange multiplier $\lambda$.
Meanwhile, we have to scale $U$ to $\frac{U}{2}$, as pointed out in Refs.
\cite{Florens2002,Florens2004}, in order to have consistent connection with the
large-$M$ limit of the O($2M$) model.
(We admit this is a bold assumption because this is based on the atomic limit and cannot explain full spectrum~\cite{Florens2002,Florens2004}.) 
Therefore, the action becomes
\begin{equation}
\begin{split}
S&=\int_{0}^{\beta} d\tau  \left\lbrace \sum_{i,\kappa, \sigma} \bar{f}_{\kappa i \sigma} \left( \partial_{\tau} -3 h\right) f_{\kappa i \sigma} + H_{t} \right. \\
& \left. + \sum_{i_{c}}  \left[ \frac{1}{2U} |\partial_{\tau} X_{i_{c}}|^2 
+ \frac{h}{2U}\left( X_{i_{c}} \partial_{\tau}  X_{i_{c}}^{*} - \text{H.c.} \right)+ \lambda |X_{i_{c}}|^2 \right] \right\rbrace \\
& + \beta N_{c}\left( 6h - \frac{h^2}{2U}-\lambda\right).
\end{split}
\end{equation}

Although the Lagrange multipliers $h$ and $\lambda$ are variables to be integrated out,
they are treated as constants of their saddle-point values in practice.
With the assumption, Green's functions of the spinon and the rotor are obtained.
The self-energies of the spinon and the rotor come from $H_{t}$ that describes the
coupling between them.
To make it simple, we simply substitute $H^{\text{MF}}_{t}=
H^{\text{MF}}_{t,f}+H^{\text{MF}}_{t,X}$ (see it in Appendix \ref{Appendix:MF}) for
$H_{t}$ in $S$, which implies that dynamical fluctuations from $H_t$ are omitted.
The Green's functions are, therefore,
\begin{align}
G_{f}^{\kappa}(i \omega_{n},\vk) =& \left[ i\omega_n - \left(\begin{array}{cc}
h_{0,\kappa}(\vk)  & h_{1,\kappa}(\vk) \\
h_{1,\kappa}^{*}(\vk) & h_{0,\kappa}(\vk)
\end{array}\right) \right]^{-1}, \\
\begin{split}
G_{X}(i\nu_{n},\vq) =& Z \beta \delta_{n,0} \delta(\vq) \\ 
 &- \frac{2U}{\left(i\nu_n+h\right)^2 - 2U\left[ \varepsilon_{X} (\vq)+\lambda \right]},
\end{split}
\end{align}
where we have absorbed $3h$ into $\mu$ in $G_{f}$ and $\frac{h^2}{2U}$ into $\lambda$ in $G_{X}$.
In order to have a stable rotor system, $\varepsilon_{X} (\vq) +\lambda \ge 0$ for all
$\vq$.
So we rewrite $\varepsilon_{X} (\vq)+\lambda$ to be $\left[\varepsilon_{X}
(\vq)-\varepsilon_{X} (\mathbf{0}) \right]+ \Delta_{\text{charge}}^2/2U$, where
$\Delta_{\text{charge}}$ stands for the Mott gap.
All the parameters $Z$, $\Delta_{\text{charge}}$, $\mu$, and correlation functions are
determined self-consistently from Eqs. (\ref{SCE1})-(\ref{SCE4}).

\bibliographystyle{apsrev4-1} % Tell bibtex which bibliography style
%\nocite{*}
\bibliography{tBG}
\end{document}